\documentclass[useAMS,usenatbib,usegraphicx]{mn2e}

\title[Water in the Near IR spectrum of Comet 8P/Tuttle]{Water in the
  Near IR spectrum of Comet 8P/Tuttle} 

\author[R.\,J. Barber et al.]{R. J. Barber$^{1}$\thanks{E-mail:
  rjb@star.ucl.ac.uk}, S. Miller$^{1}$, N. Dello Russo$^{2}$,
  M. J. Mumma$^{3}$, J. Tennyson$^{1}$, P. Guio$^{1}$\\
  $^{1}$Department of Physics \& Astronomy, University College London,
  WC1E 6BT\\ 
  $^{2}$Space Department, The Johns Hopkins University
  Applied Physics Laboratory, MD 20723-6099, USA\\ 
  $^3$Solar System Exploration Division, NASA/GSFC, Greenbelt, MD 20771, USA}

\date{Accepted XXX. Received XXX; in original form XXX}

\pagerange{\pageref{firstpage}--\pageref{lastpage}} \pubyear{2008}

\begin{document}

\label{firstpage}

\maketitle

\begin{abstract}
High resolution spectra of Comet 8P/Tuttle were obtained in the
frequency range 3449.0--3462.2 cm$^{-1}$ on 3 January 2008 UT using
CGS4 with echelle grating on UKIRT. In addition to observing solar
pumped fluorescent (SPF) lines of H$_2$O, the long integration time
(152 minutes on target) enabled eight weaker H$_2$O features to be
assigned, most of which had not previously been identified in cometary
spectra. These transitions, which are from higher energy upper states,
are similar in character to the so-called `SH' lines recorded in the
post Deep Impact spectrum of Comet Tempel~1 (Barber et al., 2007).We
have identified certain characteristics that these lines have in
common, and which in addition to helping to define this new class of
cometary line, give some clues to the physical processes involved in
their production. Finally, we derive an H$_2$O rotational temperature
of $62\pm5\,\rmn{K}$ and a water production rate of
$(1.4\pm0.3)\times10^{28}$ molecules~s$^{-1}$.
\end{abstract}

\begin{keywords}
Comet: 8P/Tuttle, line: identification, line: formation
\end{keywords}

\section{Introduction}
Cometary nuclei are the most primitive objects in the solar
system. Within about 3 AU of the sun, solar heating causes the surface
temperature of the nucleus to rise above 200 K; water ice sublimes and
other volatiles and icy dust grains are expelled from the surface. The
motion, chemistry and rate of sublimation of molecules from the icy
grains are influenced by the solar wind and solar radiation. Initially
the escaping gases form a dynamic, gravitationally-unbound atmosphere,
the coma. They are subsequently swept into the comet's plasma and dust
tails and eventually dissipate into interplanetary space. The surface
of the nucleus is processed during the course of repeated visits to
the inner solar system, becoming covered with a rubble blanket of
particles too large to be dragged away by escaping gas. During the
course of numerous perihelia, the outer surface becomes completely
depleted of volatile materials which either escape from the surface or
else migrate inwards. This creates an outer mantle of siliceous dust
(Huebner, 2008), above a layered structure in which the more volatile
species are found at greater depths (Prialnik et al., 2008). This
structure results in the great variety that is observed in active
comets.

Cometary nuclei typically have radii of $\sim$3 km and are not able to
be resolved from Earth. During close approaches, the inner coma of
some comets are resolvable with large terrestrial
instruments. However, in general, data obtained with Earth-based
instruments are indicative of conditions in a wider region of the
coma, and our understanding of conditions within the inner coma (a
region extending up to a few hundred km, where energy transfer is
collisionally-dominated, relies largely on modelling. Our
understanding of the more extended regions of the coma, where the
densities of the species are too low for thermodynamic equilibrium to
exist, also relies on modelling. In particular, models of cometary
coma predict the existence of so-called `solar pumped fluorescent'
(SPF) H$_2$O emission lines. These originate from ro-vibrational
excited upper states of the molecule, which, if observed in a
collisionally-dominated region would be characteristic of kinetic
temperatures of several thousand K, but are able to be produced in the
cold, rarefied conditions existing in cometary coma, since these
excited states, which have relatively large Einstein A coefficients,
have time to decay radiatively before they are de-excited by collision
with another molecule.

Despite the success of models in predicting SPF lines (Crovisier,
1984; Weaver and Mumma, 1984, Mumma et al., 1995), the physical
conditions existing in cometary coma are highly complex and recent
observations of Comet Tempel 1 (Barber et al., 2007) revealed a type
of emission line, not previously recorded in cometary spectra, and not
predicted by the existing cometary models. These so-called `SH' lines
originate from upper states that are excited to higher vibrational
energies than are the upper states of the SPF lines. Consequently,
their production mechanisms are not able to be explained simply, by
the existing cometary models. The near-Earth approach of Comet
8P/Tuttle in January 2008, provided an opportunity for us to search
for, assign and characterise SH emission lines in its coma. We believe
that our findings, which are presented in this paper, provide the
basis for a more thorough investigation of the physical processes
involved in the production of SH lines in cometary coma.

\section{Comet 8P/Tuttle}
8P/Tuttle is a short period comet, P$_\mathrm{orb}$ = 13.51 yr, and is
the parent of the Ursid meteor stream (Jenniskens et al., 2002). It
had been estimated that the comet has a radius of $\sim$ 7.5 km,
making it the largest of the group of 18 short period comets reported
on by Licandro et al.\,(2000). However radar images obtained on 2-4
January 2008 (Harmon et al., 2008) show a bifurcated nucleus
consisting of two lobes each 3--4 km in diameter, suggesting a
possible contact binary. Photometric measurements, have yielded
rotational times of the nucleus as being either 5.71$\pm$0.04 hr
(Schleicher and Woodney, 2007b), 7.7$\pm$0.2 hr (Harmon et al., 2008),
or multiplicities of 5.7 and 7.4--7.6~hr (Drahus et al.,
2008). However, these results may need to be re-interpreted in the
light of the probable bifurcation of the nucleus.

In the scheme proposed by Levison (1996), comets are classified by
reference to the Tisserand invariant. To a first order, this parameter
is a constant of the motion of a comet. It is based on the Jacobi
integral in the restricted three-body problem (Tisserand, 1894;
Moulton, 1947). Because 8P/Tuttle has a Tisserand invariant of less
than 2 (T$_J$ = 1.601), it is classified as a `near isotropic comet'
(NIC) of the Halley type.  However, its present orbit is not
characteristic of this class, since most NIC comets have long orbital
periods (defined as being greater than 200 yr).

8P/Tuttle also differs from most short period comets as these normally
have a Tisserand invariant 2 $\le$ T$_J $ $\le$ 3 and in the Levison
scheme are referred to as `ecliptic comets' (EC).

Being a short period NIC, the chemistry of 8P/Tuttle is of particular
interest. Unlike most short period comets, which are of the EC type,
8P/Tuttle is believed to have been formed in the region of the giant
planets. It therefore has the potential to reveal possible differences
between EC and NIC objects. Because it has made many approaches to the
Sun, 8P/Tuttle's surface may be highly processed like that of an EC. If
then, despite having a highly-processed surface, its spectra reveal a
chemistry that is more like that of NICs than of ECs, it may be
possible to identify differences in the chemistries of various regions
of the very early solar system.

Even though, with the exception of the 1953 approach, 8P/Tuttle had
been seen on each orbit since its discovery in 1858, until recently
there had been no detailed investigation of its near-IR spectrum.
During the 2008 apparition, 8P/Tuttle approached to within 0.25 AU of
the Earth and was favourably placed for observing (see for example,
Bonev et al., 2008 and B\"{o}hnhardt et al., 2008).

\section{Water in Cometary Spectra}\label{Cometary_spectra}
In the near IR, the spectra of comets are rich in ro-vibrational
H$_2$O emission lines which can, in principle, be observed from
Earth. However, because of the low temperatures that characterise
cometary comae, the strongest water lines are fundamental transitions
(that is to say, transitions to ground vibrational states) from
low-energy ro-vibrational states. Photons from these transitions
are absorbed by water vapour in the Earth's atmosphere, the molecules
of which are in ground vibrational states.

 Close to the nucleus the density is sufficiently great (generally
 upward of 10$^6$ molecules cm$^{-3}$) to ensure that the molecules
 are collisionally thermalised in low {\it J} ground vibrational
 states, with a rotational temperature equal to the local kinetic
 temperature (Weaver and Mumma, 1984). There is no precise boundary
 between the collisionally-dominated and the less dense, radiatively
 dominated region, as the breakdown of thermal equilibrium occurs at
 different distances from the nucleus for each of the allowed
 rotational transitions between ground vibrational states. These
 distances vary between several tens of kilometres to several
 thousands of kilometres (Bockel\'{e}e-Morvan, 1987). Moreover, beyond
 the region where H$_2$O--H$_2$O collisions are able, alone, to
 maintain local thermodynamic equilibrium (LTE) there is a region
 where the additional contribution of H$_2$O--electron collisions and
 line-trapping effects maintains LTE to greater radial distances
 (Bockel\'{e}e-Morvan, 1987; Zakharov et al., 2007). In particular,
 Xie~and~Mumma (1992) showed that at distances of several thousand
 kilometres, e$^-$--H$_2$O collisions play an important role in
 exciting water molecules. In Section\,\ref{Origins} we discuss a
 number of mechanisms, including e$^-$--H$_2$O collisions, that may
 possibly be involved in the formation of the so-called `SH' water
 lines that were observed to be present in our 3 January 2008 UT
 spectrum of 8P/Tuttle. SH, or `stochastic heating' lines were first
 identified in the `Deep Impact' spectrum of Comet Tempel~1 (Barber et
 al., 2007), and were subsequently noted to be present in another
 spectrum of the same comet (Mumma et al., 2005, Figure~2). Their
 characteristics are discussed in Section\,\ref{Data_analysis}.

 At distances greater than a few thousand kilometres from the nucleus
 the density of the species is sufficiently low, and the mean free
 path between collisions large enough, for `fluorescence equilibrium'
 to exist. That is to say, radiation is due to the balance between
 solar pumping to vibrationally-excited states and subsequent
 spontaneous radiative decay (Crovisier, 1984). Under
 collisionally-thermalised conditions, these upper states would
 typically require temperatures of several thousand Kelvin for them to
 be sufficiently populated for their collisionally-induced radiation
 to lower states to be observable. The observation in cometary spectra
 of transitions from vibrationally-excited upper states, indicates
 that they must be produced in regions where Boltzmann statistics do
 not apply.

When vibrationally excited states decay radiatively to lower energy
ro-vibrational states above the ground vibrational state (hot bands),
the photons emitted are not absorbed in the Earth's
atmosphere. Recognition of this fact was a key factor in developing
the `solar-pumped fluorescent' (SPF) approach to analysing cometary
spectra (Mumma et al., 1995).  Spectra containing these SPF
transitions enable information to be derived about the rotational
temperature, T$_\mathrm{rot}$ of the inner coma and the water
production rate, Q$_\mathrm{H_{2}O}$. A comparison of this latter
value with the production rates of trace species provides the
abundance ratios that are fundamental to understanding cometary
chemistry.

Because the upper states populated by solar pumping are not
collisionally thermalised, the derivation of physical quantities from
line intensities requires knowledge of all the possible excitation as
well as the alternative de-excitation routes. Each upper excited
ro-vibrational state can be populated directly by radiative pumping of
rotational levels of the ground vibrational state, as well as by
cascade from higher energy states (Crovisier, 1984). The relative
probabilities of these various upward and downward routes are
temperature-dependent.  Models have been developed to compute the {\it
g}-factors for each SPF line, from which line intensities are able to
be calculated as a function of temperature (Crovisier, 1984; Mumma et
al., 1995, Dello Russo et al., 2000). These models require a knowledge
of the Einstein\,A coefficients for the downward transitions. Ideally,
the upward pumping rates should be computed using the Einstein B
coefficients. However, for simplicity, the temperature-dependent
vibrational band intensities are frequently used to compute the upward
pumping rates. For example, Dello Russo et al.\, (2004) used
Einstein\,A coefficients from the BT2 synthetic water line list
(Barber et al., 2006) to model {\it g}-factors for all the H$_2$O SPF
transitions in the: $\nu$$_1$+$\nu$$_3$--$\nu$$_1$,
$\nu$$_1$+$\nu$$_3$--$\nu$$_3$, 2$\nu$$_1$--$\nu$$_1$,
2$\nu$$_1$--$\nu$$_3$ and
$\nu$$_1$+$\nu$$_2$+$\nu$$_3$--$\nu$$_1$--$\nu$$_2$ vibrational bands,
from upper states having {\it J} $\le$ 7.

The 2.9--3.0 $\mu$m (3300--3450 cm$^{-1}$) spectral region is particularly
rich in SPF lines and is also largely devoid of other molecular
species. The availability of accurate {\it g}-factors has enabled this
region to be used in determining cometary rotational temperatures and
water production rates (e.g. Dello Russo et al., 2004 and 2006).

\section{Observations}
We observed 8P/Tuttle on its approach to the Sun (perihelion being on
27 January UT), on the nights of 3, 4 and 5 January 2008 UT from
UKIRT, Hawaii, using the echelle grating on CGS4. We report here on the
spectra that we obtained on 3 January UT, in the frequency range
3\,440.6--3\,462.6 cm$^{-1}$. The heliocentric distance on this date
was $\sim$1.09 AU, and the geocentric distance and velocity were
$\sim$0.25 AU and $\sim$3.3 km s$^{-1}$ respectively; the latter
figure corresponds to a red shift of 0.038 cm$^{-1}$, which was
equivalent to 1 pixel on the array, and less than the minimum
resolution of our instrument, which at R = 37\,000 was 0.093 cm$^{-1}$
at 3\,450 cm$^{-1}$. The total visual magnitude of 8P/Tuttle at the
time of our observations was $\sim$5.8. However our observations were
of the bright inner coma region which presented a point source in our
instrument, and required long integration times.

We used a slit of 2-pixel width of length 90 arcsec, oriented
east-west. The scale was 0.86 arcsec per pixel in the spatial
direction (156 km at the comet) and 0.41 arcsec in the spectral
direction, (149 km at the comet for a 2-pixel slit). The telescope was
dithered to give 2x2 sampling every 40 seconds. Observations were
acquired using a standard ABBA sequence, which was achieved by nodding
the telescope 22 pixels along the slit. In this mode, the comet signal
was present in both the A and B beams (rows 114 and 92 respectively),
which, compared to nodding to blank sky, increased the signal to noise
ratio by a factor of $\sqrt{2}$, when both signals were added using
the (A-B)-(B-A) routine to remove the sky. Each ABBA series
represented a total of 160 seconds integration time. Total time
on-target was 152 minutes. Spectra of an A0V star, HD 6457, were
obtained at the beginning and end of the observing session. Standard
star spectra are required in order to adjust for frequency-dependent
atmospheric transmission, and also to enable absolute flux
calibration.  Before dividing by the standard star, we removed high
frequency noise from both the 8P/Tuttle and standard star data using
Fourier transform smoothing. We rejected all data at frequencies where
the atmospheric transmission was less than 20\%, and this resulted in
19\% of the data in the 3\,440.6 to 3\,462.6 cm$^{-1}$ range being
excluded. We noted that transmission was poor at frequencies less than
3449.0 cm$^{-1}$ and above 3\,460.2 cm$^{-1}$. We therefore decided to
limit the spectral region under investigation to that lying between
these frequencies. In this reduced, 11.2 cm$^{-1}$, frequency range
less than 3\% of the data failed to meet our 20\% minimum atmospheric
transmission requirement, and the average transmission in this reduced
frequency range was 66\%. After dividing by the standard star, flux
calibration was achieved by multiplying the result by the
frequency-dependent flux of HD 6457, calculated relative to Vega as a
photometric 0.0 magnitude standard (Hayes, 1985), and assuming a black
body function with {\it {T}}$_\mathrm{eff}$~=~9\,480 K (Tokunaga,
2000, p151).

Because of the lack of arc lines in this spectral region, we
frequency-calibrated our data using the position of several known SPF
lines that were present in the 8P/Tuttle spectrum. This had the effect
of transforming the observed spectrum into the rest frame. The
fully-reduced, flux-calibrated spectrum is given in Figure~1.

\begin{figure*}
\begin{center}

\label{spectrum}

  \includegraphics[width=1.0\textwidth]{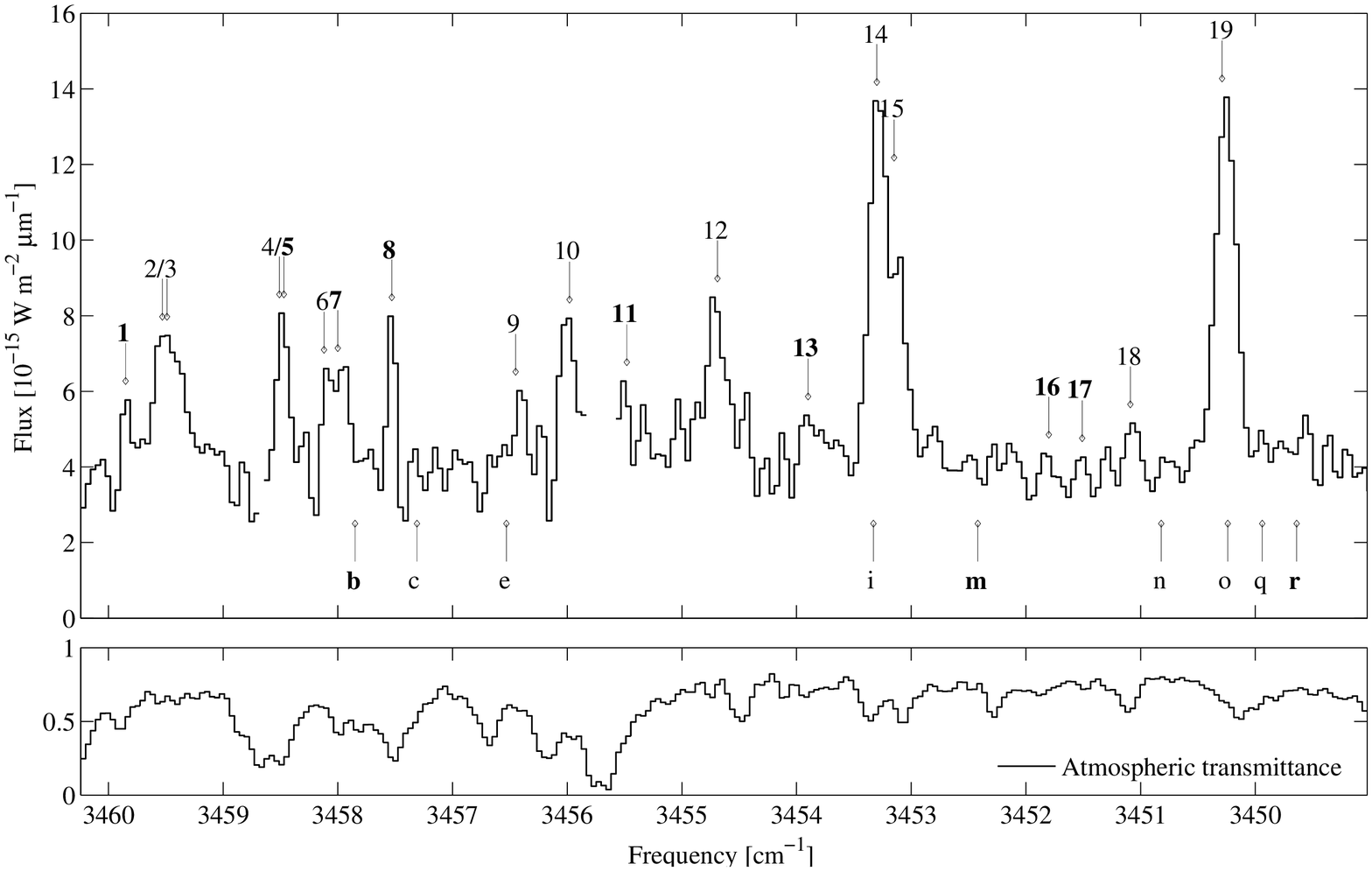}

\caption{The upper chart shows the spectrum of Comet 8P/Tuttle, UT 3
  January 2008 adjusted to rest frame and plotted after dividing by
  standard star HD 6457 (normalised) to adjust for frequency-dependent
  atmospheric transmission. The spectra of both the comet and standard
  star have been Fourier Transform smoothed to remove high frequency
  noise. Atmospheric transmission is shown in the lower chart, and no
  spectral data is displayed where transmission is less than 0.20. The
  figures above the spectrum indicate lines that have been assigned in
  Table\,\ref{assignments_table}, the numbering above the plot
  corresponds to that given in each entry in this table. Bold font is
  used to indicate SH transitions; SPF lines are numbered using normal
  weight font. The letters below the plot correspond to the nine SH
  lines which we would expect to be present in the spectrum (see
  text), but which were not positively identified. Their numbers
  correspond with those given in the first column of each entry in
  Table\,\ref{SH_table}. Six of these either correspond to weak
  feature (S/N $\leq$\ 2), or are at frequencies where they would be
  undetectable due to blending with stronger lines. Of the twenty-six
  SH lines listed in Table\,\ref{SH_table}, we are unable to explain
  the absence of three, and these are identified with bold letters.}
\end{center}
\end{figure*}

\section{Data Analysis}\label{Data_analysis}
We compared the features in Figure~1 with our BT2 database of H$_2$O
SPF line positions and {\it g}-factors. Nine of these features are at
frequencies corresponding to known SPF lines (there are actually eleven
SPF lines, with two features being blends of two SPF lines). Details
are given in Table\,\ref{assignments_table}. However, Figure~1 also
includes a number of other features, whose shape and intensity above
the continuum (S/N $\geq$2) suggest that they are spectral lines,
rather than noise, but whose frequencies do not correspond to SPF
lines.

In order to assign these other features, we employed the methodology
detailed in Barber et al.\,(2007) for the identification of non-SPF
features in the post Deep Impact spectrum of Comet Tempel~1. Initially
we used the BT2 line list to generate synthetic emission spectra for water at a
temperature of 3\,000 K in the 3\,449.0--3\,462.2 cm$^{-1}$ frequency
range, restricting the value of {\it J} for the upper state to 8, and
the intensity of the weakest line to be generated was limited to
10$^{-3}$ of the strongest. The output, which consisted of
about 600 individual lines was convolved to match the resolving
power of the CGS4 echelle, R=37\,000. 

Some of the features in this synthetic spectrum are at frequencies
corresponding to features in the spectrum of 8P/Tuttle. However, the
relative intensities of the features in the observed and synthetic
3\,000 K spectra do not agree. The reason is that the observed
emission lines originate in low-density regions where
vibrationally-excited upper states have sufficient time to decay
radiatively before they can be collisionally de-excited. These upper
states are populated by pumping low-lying rotational states and by
cascade from higher levels (of which there will be many, most of which
contribute little and can therefore be disregarded). Since the water
molecules are not collisionally thermalised, the populations of the
upper states can be substantially greater than would be the case if
Boltzmann statistics applied, and the frequencies and intensities of
the emissions from these states are frequently similar to those of
emission lines from H$_2$O vapour in LTE regions at temperatures of,
say, 3\,000 K.

Our synthetic spectrum contains many more features than are present in
the spectrum of 8P/Tuttle. It is therefore important to try to
identify specific characteristics of the observed lines that
differentiate them from lines in the synthetic spectrum that are
absent in the observed spectrum.

When synthetic spectra are generated using BT2, one of the output
files gives the frequency, assignment, intensity, Einstein A
coefficient and lower state energy of every water transition in the
frequency range, down to a specified intensity cut-off. Using these
data, we were able to identify certain characteristics of those lines
in our synthetic spectra whose positions matched those of features in
the observed spectrum, that differentiated them from those that did
not. This information enabled us to constrain the parameters applied
when generating the BT2 synthetic spectrum, which reduced the number
of lines in the synthetic spectrum and hence the risk of confusing
noise and real spectral lines.

Before detailing these tighter parameters and our findings, we comment
briefly on the ro-vibrational states of water. The three vibrational
modes of the H$_2$O molecule, $\nu_1, \nu_2$ and $\nu_3$ (symmetric
stretch, bend and asymmetric stretch respectively), require different
amounts of energy for one quanta of excitation. Also, the magnitude
of each increment decreases as the total internal energy of the
excited molecule increases. In the energy range in which we are
interested, each quanta of symmetric and asymmetric stretch represents
$\sim$3\,400-3\,700 cm$^{-1}$. These values are approximately twice
the energy of one quanta of bend. Hence it is convenient to talk in
terms of `polyads', where each polyad, designated $\nu$ is the
equivalent of one quanta of stretch or two of bend. Single bend quanta
are denoted by~`$\delta$'.

We noted that all the SPF lines in the observed spectrum came from the
2$\nu$ polyad, whilst the large majority of non-SPF lines that matched
features in our synthetic spectra, were from the 3$\nu$ and
3$\nu$+$\delta$ polyads.  We also noted that none of the upper states
had more than two $\nu_{\mathrm2}$ quanta. In addition, the lines in
our synthetic spectrum that corresponded to non-SPF features in
Figure~1 all had Einstein A coefficients $\ge$ 1.0 s$^{-1}$, and were
on average higher than those of the SPF lines, the average A$_{if}$
being 23.0 and 7.4~s$^{-1}$ respectively. The SPF lines had upper
states with energies in a very narrow range: 7\,242--7\,613 cm$^{-1}$,
whilst the non-SPF lines had upper states with energies in the range
10\,365 --12\,940 cm$^{-1}$.

Rotational excitation, also affects which lines are observed in
cometary spectra. This is defined in terms of the asymmetric top
labels, $J,\ K_{\mathrm{a}},\ K_{\mathrm{c}}$ which represent total
angular momentum and the prolate and oblate levels respectively. We
designate ro-vibrational states ($\nu_1\ \nu_2 \nu_3$)[$J\
K_{\mathrm{a}}\ K_{\mathrm{c}}$] and when assigning transitions, give
the upper state first.

An examination of the $\sim$ 600 transitions in our synthetic spectrum
revealed that none of the features in the spectrum of 8P/Tuttle
corresponded to lines whose upper state had {\it J}~$\ge$ 5. We also
noted that none of the lines in our observed spectrum,assigned in
Table\,\ref{assignments_table}, correspond to very weak features in
our synthetic spectrum. They all correspond to lines in our synthetic
spectrum having intensities greater than $\sim$0.5\% of the intensity
of the strongest line in the synthetic spectrum. In all of these
respects the non-SPF lines that we identified in the spectrum of
8P/Tuttle had the same characteristics as the SH lines observed in the
`Deep Impact' spectrum of Comet Tempel 1 (Barber et al., 2007).

This information provided us with with a second, even tighter, set of
parameters with which to search for SH lines. We produced a second
synthetic BT2 spectrum with the constraints: {\it J} $\le$ 5, 3$\nu$
or 3$\nu$+$\delta$ polyad, A$_{if}$ $\ge$ 1.0 s$^{-1}$ and a minimum
intensity cut-off equivalent to 0.5\% of the intensity of the
strongest line in the frequency range. Also we lowered the temperature
of the synthetic spectrum to 2\,500 K, which further reduced the total amount
of data generated.

Although it still includes all the SPF lines that were present in our
observed spectrum, this second synthetic LTE spectrum contains only 26
possible SH lines. Using these data, we were able to assign seven
unblended SH lines in the observed spectrum.  These are detailed in
Table\,\ref{assignments_table}, which also gives E$_\mathrm{upper}$,
A$_{if}$; the type of transition, SPF or SH; the nuclear spin species
identity (ortho/para) and the estimated signal-to-noise ratio of the
lines.  In addition to the unblended SH lines, we identified a feature
centred at 3458.49 cm$^{-1}$. There is a known SPF line,
(200)[423]-(100)[432], at 3458.51 cm$^{-1}$. However, at 62 K, the
{\it g}-factor for this line is only 8.2$\times10^{-10}$, which
equates to a line of only $\sim$3\% of the intensity of the feature
observed at 3\,458.49 cm$^{-1}$. The BT2 synthetic spectrum contains
only one other water line in the 3\,458.49 cm$^{-1}$ region, an SH
line, (022)[432]-(120)[541] at 3458.47 cm$^{-1}$. We therefore
conclude that this line is also present in the observed spectrum of
Comet 8P/Tuttle, which represents an eighth SH detection.

Barber \textit{et al.}\,(2007) assigned two unblended SH transitions
in the spectrum of Comet Tempel~1 at 3\,453.39 and 3\,451.51 cm$^{-1}$
and a feature at 3\,453.90 cm$^{-1}$ which they suggest might be a
blend of two lines: (211)[322]--(210)[221], which has a
3$\nu$+$\delta$ polyad upper state, of energy 12\,354 cm$^{-1}$ and
(103)[110]--(102)[110], which has a 4$\nu$ polyad upper state of
energy 14\,356.8 cm$^{-1}$. The spectrum of 8P/Tuttle in Figure~1 also
contains a feature at 3\,453.90 cm$^{-1}$, and based on the polyad
constraints that we detail above, we are inclined to believe that this
is the unblended (211)[322]--(210)[221] line, and that there is no
observable (103)[110]--(102)[110] component. This view is reinforced
by considerations of $K_{\mathrm {a}}$ and $K_{\mathrm {c}}$ (see
below). Figure~1 also contains a feature (S/N = 2.1) at 3\,451.51
cm$^{-1}$, which we identify as the (220)[212]-(021)[111] SH
transition, which is in agreement with Barber \textit{et al}'s
assignment. However, there is no feature in our observed 8P/Tuttle
spectrum at 3\,453.39 cm$^{-1}$. This may be because of the width of
the strong feature (a blend of two SPF lines) centred at about
3\,453.25 cm$^{-1}$, since a line at 3\,453.39 cm$^{-1}$ would be in wing
of this feature. It should also be noted that the line that Barber
\textit{et al} identified in Tempel~1 at 3\,453.39 cm$^{-1}$ is the
(210)[101]-(011)[110] transition, whose upper state is of the
2$\nu$+$\delta$ polyad, at an energy of 8\,784.7
cm$^{-1}$, and hence would not satisfy the 3$\nu$ and 3$\nu$+$\delta$
test that we are applying for SH lines in the current paper.

Having identified that the observed spectrum of 8P/Tuttle contains
eight of the 26 SH lines appearing in Table\,\ref{SH_table}, we
examined the data for additional characteristics that might
distinguish the observed eight from the unobserved 18, and noted one
further feature that is common to all the observed SH lines; they all
have $K_{\mathrm{a}}$ $\geq$ $J$-1. Now of the 18 lines in Table\,3
that were not observed, half had $K_{\mathrm{a}}$ $\geq$ $J$-1, and
half did not. Moreover, of the nine lines in Table\,\ref{SH_table}
that had $K_{\mathrm{a}}$ $\geq$ $J$-1, but which were not identified
in the observed spectrum, only the (300)[542]-(101)[441] transition at
3\,457.85 cm$^{-1}$, labelled \textbf{\textit{b}} in
Table\,\ref{SH_table} was definitely absent. Adopting a cautious
approach we also include the (220)[322]-(120)[413] transition at
3\,452.42 cm$^{-1}$ and the (003)[440]-(002)[541] transition at
3\,449.64 labelled \textbf{\textit{m}} and \textbf{\textit{r}}
respectively in Table\,\ref{SH_table} as being not present even though
there is a weak feature in Figure\,1 centred at 3\,452.48 cm$^{-1}$,
which is only 0.06 cm$^{-1}$ away from \textbf{\textit{m}}, and
therefore theoretically unresolvable from it, and a 2.1 S/N feature
centered at 3\,449.55, which would also theoretically be unresolvable
from the synthetic SH line \textbf{\textit{r}}. Of the remaining six
lines in Table\,\ref{SH_table} that have $K_{\mathrm{a}}$ $\geq$
$J$-1, but were not observed to be present in Figure\,1, two are at
frequencies where they would be blended with stronger SPF features,
and four are at frequencies corresponding to weak features (S/N $\leq$
2).  On this basis, we estimate the probability as 1 in 763 that the
property common to all SH lines assigned here ($K_{\mathrm{a}}\ \geq\
J$-1) is a chance event. We therefore conclude that this
characteristic has a physical cause, most probably relating to the
mechanism by which SH lines are produced. We discuss this in more
detail in Section\,\ref{Origins}, but comment here that for a given
$J$, there are 2$J$+1 possible combinations of $K_{\mathrm{a}}$ and
$K_{\mathrm{c}}$, and the higher the value of $K_{\mathrm{a}}$, the
higher is the energy of the state. The preferential population of
higher value $K_{\mathrm{a}}$ states is the reverse of what would be
observed in a Boltzmann distribution, and this inversion of states
strongly suggests that the upper levels of SH transitions are
populated by cascade from more energetic levels, rather than by
pumping from ground vibrational states, which is the principal method
of populating the upper states of SPF transitions.

In addition to analysing our own 8P/Tuttle data, we examined the
spectra of comet C/1999 H1 Lee obtained on 19 and 21 August 1999 using
NIRSPEC on Keck (Dello Russo et al., 2006). These contained several
weak, unassigned, features. We noted that the frequencies of three of
these features corresponded to SH lines in the spectrum of
8P/Tuttle. These are:\\ (022)[431]--(120)[542] at 3\,459.85
cm$^{-1}$\\ (211)[322]--(210)[211] at 3\,453.90 cm$^{-1}$\\
(220)[212]--(021)[111] at 3\,451.51 cm$^{-1}$.\\ We do not claim that
our assignments are totally secure. In particular we observe that the
feature at 3\,451.51 cm$^{-1}$ was only present in the spectrum
obtained on 19 August, whilst a line at 3\,453.90 cm$^{-1}$ would be
likely to appear blended (see below). However, the fact that these two
lines were also observed in Comet Tempel~1 increases our confidence in
these assignments. The third line, at 3\,459.85 cm$^{-1}$ is outside
the frequency range examined in Barber et al.\,(2007). Dello Russo et
al.  state that the S/N ratio of the feature centred at
$\sim$3\,453.85 cm$^{-1}$ in their spectra of comet Lee is 7.9, and
indicate that its position corresponds to the (110)[313]--(010)[422]
SPF line at 3\,453.88 cm$^{-1}$. However, the $g$ factor for this line
is too low to account for a signal of this intensity, which leads us
to believe that the main component in the blend is likely to be the
(211)[322]--(210)[211] SH line. Comparing the intensities of the
features at 3\,459.85 and 3\,451.51 cm$^{-1}$ on the two nights with
that at 3\,453.85 cm$^{-1}$, we infer that the signal-to-noise
ratios of the SH lines alone are between 2.5 and 3.5.

Using the original data file and applying Fourier Transform smoothing
to remove high frequency noise, we identified several water lines that
were unassigned in the high resolution spectrum of 8P/Tuttle obtained
at ESO with CRIRES on 27 January 2008 UT, when the heliocentric
distance was 1.03 AU (B\"{o}hnhardt et al., 2008, Fig. 1C). These
include SPF lines at 3\,451.09, 3\,456.45 and 3\,458.12 cm$^{-1}$, all
of which are also present in Table\,\ref{assignments_table}. We also
noted that the spectrum contained an unassigned feature at 3\,455.48
cm$^{-1}$ which is the frequency of the (013)[330]--(012)[413] SH line
in Table\, \ref{assignments_table}. We estimate the S/N ratio of this
feature to be $\sim$3 and note that its width suggests that it may be
a blend of more than one line.
 
Finally, we note that the unassigned feature at 3\,453.90 cm$^{-1}$ in
the spectrum of Comet 73P/Schwassmann-Wachmann 3C obtained by
N.\,Dello Russo on 15 May 2006 UT (private communication), corresponds
to the (211)[322]--(210)[211] SH line in Table\,\ref{assignments_table}.

On the basis of our identification of SH lines in comets of different
types, we conclude that this class of water lines, which until
recently had not been identified in cometary spectra, is probably
always present at heliocentric distances of up to $\sim$1.5
AU. However, it is noted that the SH lines observed vary from
comet to comet, and indeed between spectra of the same comet obtained
on different dates. Table\,\ref{assignments} contains details of the
four comets in which SH lines have been recorded.

\begin{table}
\caption{List of comets in which SH water lines have been observed
\label{assignments}}
\smallskip
\begin{tabular}{rrccrr}
\hline
Name & P (yr) & Type & Origin & r$_\mathrm{h}$ & T$_\mathrm{rot}$ K\\[0.5ex]
\hline
 9P/Tempel 1   & 5.5     & $^1$EC & Kuiper & 1.51 & $\sim$40 \\
C/1999 H1(Lee) & 78\,000 &     NIC & Oort   & 1.06 & $\sim$78 \\
73P/SW3        & 5.4     & $^2$EC & Kuiper & 1.03 & $\sim$110 \\
8P/Tuttle      & 13.5    & $^3$NIC & Oort   & 1.09 & $\sim$62  \\

\hline
\multicolumn{4}{l}{EC: Ecliptic Comet}\\
\multicolumn{4}{l}{NIC: Near Isotropic Comet}\\
\multicolumn{4}{l}{$^1$ artificial impact}\\ 
\multicolumn{4}{l}{$^2$ fragmented}\\
\multicolumn{4}{l}{$^3$ bifurcated nucleus}\\
		
\hline
\end{tabular}
\end{table}  

\section{Rotational temperature and H$_2$O production rate}
The procedure that we used to determine the temperature of the inner
coma is based on the fact that the {\it g}-factors of some SPF
transitions can behave quite differently as a function of temperature:
some {\it g}-factors increase with temperature, whilst others decrease
(Dello Russo et al., 2004). In broad terms, SPF lines whose upper states have
been pumped from higher {\it J} values become more intense as
temperature rises, whilst those pumped from the lower {\it J} states
show little increase in intensity at higher temperatures, and in
the case of those pumped from the lowest rotational states, may
actually weaken with increasing temperature.

Using the {\it g}-factors for the SPF lines in our spectral frequency
range (Dello Russo et al., 2004), we generated SPF synthetic spectra,
convolved to the resolving power of our instrument. The spectra were
produced using the Fortran program, spectra-BT2 (Barber et al., 2006)
which is available in electronic form via: {\it
http://www.tampa.phys.ucl.ac.uk/ftp/astrodata/water/BT2} and were
generated at various temperatures with 5 K increments (the smallest
temperature difference that produced measurable differences in the
synthetic spectra). The spectral region contains many SPF transitions
in the five vibrational bands, for which {\it g}-factors were computed
by Dello Russo. Some of these lines are not detectable at low
rotational temperatures, whilst others are at frequencies that fall
within regions of low atmospheric transmission. We assigned nine SPF
lines in Figure\,1, of which two are blended. With the exception of
the SPF/SH blend at 3458.49 cm$^{-1}$ (which as discussed above is
predominantly an SH line) we used all of these features for
temperature diagnostics.

On comparing the relative intensities of the synthetic spectra with
our observed spectrum, normalised to the cometary continuum level, we
observed that there was excellent agreement between the relative
intensities of all the SPF lines at 62 $\pm$5 K, and consequently we
conclude that this was the rotational temperature of the inner coma of
8P/Tuttle at UT 3.30 January 2008 (the mid-point of our observations)
when the geocentric distance was $\sim$0.25 AU. In arriving at this
temperature, we have assumed the normal ortho-para ratio (OPR), which
is 3:1. Some comets have been observed to have sub-normal
OPRs. However, as all seven features used to derive the temperature
of the inner coma are transitions between ortho states, our results
are not sensitive to the OPR. Our derived rotational temperature is
consistent with those obtained by Bonev et al. on 22 and 23 December
2007, of 60$\pm$15 K and 50$\pm$10 K respectively, when the geocentric
distances were $\sim$0.32 and $\sim$0.31 AU respectively.

\begin{table*}
\caption{Assignments of SPF and SH lines in the post-impact spectrum
of 8P/Tuttle. In order, the columns give: a reference number: the rest
frequency in cm$^{-1}$; the assignment, upper level given first
(vibrational quantum numbers in round brackets, rotational quantum
numbers, $J,\ K_{\mathrm{a}},\ K_{\mathrm{c}}$ in square brackets);
energy of the upper level in cm$^{-1}$; Einstein A coefficient from
BT2; type of line, SPF/SH; ortho/para state; S/N ratio and comments
\label{assignments_table}}
\smallskip
\begin{tabular}{rrcrrrcrl}
\hline
Ref. & Freq. & Identification & E$_\mathrm{upper}$ & A$_{if}$ & Type &
O/P & S/N & \hspace{5em}Comment\\
 & cm$^{-1}$ & see text & cm$^{-1}$ & s$^{-1}$ &  &  & &\\[0.5ex]
\hline
1 & 3459.85 & (022)[431]-(120)[542] & 10\,934 &  3.2 & SH & P  & 3.0 &
Also seen in Lee$^{\dagger}$ \\
2 & 3459.53 & (101)[111]-(001)[202] & 7\,285 &  1.1 & SPF & O &
4.2 & Possibly blended with 3, and \textit{a} in Table\,3\\
3 & 3459.49 & (101)[431]-(100)[532] & 7\,613 & 22.2 & SPF & O &
4.2 & Possibly blended with 2 and \textit{a} in Table\,3\\
4 & 3458.51 & (200)[423]-(100)[432] & 7\,489 &  1.1 & SPF & O & 5.1
& Intensity suggests blending with 5\\
5 & 3458.47 & (022)[432]-(120)[541] & 10\,933 & 3.1 & SH & O & 6.1
& Blended with 4, see above comment \\
6 & 3458.12 & (101)[000]-(001)[111] & 7\,250 &  3.6 & SPF & O & 4.7 &\\
7 & 3458.00 & (300)[541]-(101)[440] & 11\,166 & 4.9 & SH & O & 4.7 &\\
8 & 3457.53 & (003)[432]-(002)[533] & 11\,382 & 71.2 &  SH & P & 4.3 &\\
9 & 3456.45 & (101)[422]-(100)[523] & 7\,552 & 31.9 & SPF & O & 2.7 &\\
10 & 3455.98 & (200)[303]-(100)[414] & 7\,334 &  4.5 & SPF & O &
5.5 & Possibly blended with \textit{f} in Table\,3\\
11 & 3455.48 & (013)[330]-(012)[431] & 12\,940 & 51.3 &SH & P & 3.5 &
Also seen in 8P/Tuttle$^{\ddagger}$\\
12 & 3454.69 & (101)[211]-(001)[220] & 7\,342 &  1.8 & SPF & O & 5.9 &\\
13 & 3453.90 & (211)[322]-(210)[221] & 12\,354 &  8.5 & SH & O &
3.0 & Also seen in Lee$^{\dagger}$, 8P/Tuttle$^{\ddagger}$, Tempel\,1\,$^{\star}$\\
14 & 3453.30 & (200)[110]-(100)[221] & 7\,242 &  4.7 & SPF & O &
13.3 & Possibly blended with \textit{i} in Table\,3\\
15 & 3453.15 & (101)[202]-(100)[321] & 7\,318 & 1.7 & SPF & O & 7.4 &\\
16 & 3451.80 & (121)[322]-(120)[423] & 10\,550 & 37.2 & SH & O & 2.1 &\\
17 & 3451.51 & (220)[212]-(021)[111] & 10\,365 & 4.4 & SH & O & 2.1 &
Also seen in Lee$^{\dagger}$ 8P/Tuttle$^{\ddagger\otimes}$\,Tempel\,1$^{\star}$ \\
18 & 3451.09 & (101)[413]-(001)[422] & 7\,517 &  1.5 & SPF & O & 2.3 &\\
19 & 3450.29 & (200)[110]-(001)[111]& 7\,242 &  6.6 & SPF & O & 14.6 &
Possibly blended with \textit{o,p} in Table\,3\\
\\
\multicolumn{9}{l}{$^{\dagger}$ Dello Russo \textit{et al}., 2006 \hspace{1em}
$^{\ddagger}$ B\"{o}hnhardt et al.,\,2008 \hspace{1em} $^{\star}$
  Barber \textit{et al.},\,2007 \hspace{1em}$^{\otimes}$ marginal
  detection only}\\
\hline
\end{tabular}

\end{table*}

\begin{table*}
\caption{List of all the SH lines in a BT2 synthetic emission spectrum
  at 2\,500 K in the frequency range 3\,449.0 to 3\,460.2 cm$^{-1}$,
  satisfying the empirical sort parameters detailed in the text. The
  standard convention is used for assigning the transitions (see
  Table\,2). Lines that have been observed carry the same reference
  number as in Table\,2, whilst those that are not observed in
  Figure~1 are denoted by a lower case letter in the first column. It
  is noted that no observed line had $J{-}K_{a}{\leq1}$. Applying this
  additional sort parameter, we conclude that there are a maximum of
  three SH lines which we might expect to be observed in this
  frequency region that are able to be identified as not being present
  in Figure~1.  Units of intensity are Wm$^{-3}$cm mol$^{-1}$.
\label{SH_table}}
\smallskip
\begin{tabular}{rrcrrrrcl}
\hline
Ref. & Freq. & Identification & E$_\mathrm{upper}$ & O/P & A$_{if}$ &
I$_{2500 K}$ & $J{-}K_{a}{\leq1}$ & \hspace{3em}Comment\\
& cm$^{-1}$ & see text & cm$^{-1}$ & & s$^{-1}$ & & &\\[0.5ex]
\hline
1 & 3459.85 & (022)[431]-(120)[542] & 10\,934 & P & 3.2 & 3.1E-19 &
Yes & Seen\\	
a & 3459.50 & (201)[514]-(101)[423] & 10\,996 & P & 5.9 & 6.7E-19 & & Would be blended with 2 in Table\,2\\
5 & 3458.47 & (022)[432]-(120)[541] & 10\,933 & O & 3.1 & 8.9E-19 &
Yes & Seen\\
7 & 3458.00 & (300)[541]-(101)[440] & 11\,166 & O & 4.9 & 1.5E-18 &
Yes & Seen\\
b & 3457.85 & (300)[542]-(101)[441] & 11\,165 & P & 4.9 & 5.0E-19 & Yes &
Not seen and $J{-}K_{a}{\leq1}$\\
8 & 3457.53 & (003)[432]-(002)[533] & 11\,382 & P & 71.2 & 5.3E-18 &
Yes & Seen\\
c & 3457.31 & (211)[321]-(210)[220] & 12\,360 & P & 14.3 & 4.7E-19 &
Yes & Coincides with a weak feature in Fig.~1\\
d & 3457.28 & (220)[202]-(021)[101] & 10\,352 & P & 6.6 & 4.9E-19 & &\\
e & 3456.53 & (013)[321]-(012)[422] & 12\,772 & P & 93.6 & 2.4E-18 &
Yes & Coincides with a weak feature in Fig.~1 (blend)\\
f & 3455.96 & (201)[202]-(200)[101] & 10\,681 & O & 20.4 & 3.8E-18 & &
Would be blended with 10 in Fig.~1\\
11 & 3455.48 & (013)[330]-(012)[431] & 12\,840 & P & 51.3 & 1.3E-18 &
Yes & Seen\\
13 & 3453.90 & (211)[322]-(210)[221] & 12\,354 & O & 8.5 & 8.4E-19 &
Yes & Seen\\
g & 3453.51 & (310)[523]-(111)[422] & 12\,575 & O & 15.0 & 2.0E-18 & &\\
h & 3453.45 & (003)[422]-(002)[523] & 11\,332 & O & 84.7 & 1.9E-17 & &\\
i & 3453.33 & (003)[431]-(002)[532] & 11\,384 & O & 73.2 & 1.6E-17 &
Yes &Would be blended with 14 and 15 in Fig.~1\\
j & 3452.90 & (003)[202]-(002)[321] & 11\,100 & O & 4.7 & 6.8E-19 & &\\
k & 3452.63 & (211)[303]-(210)[202] & 12\,282 & P & 24.4 & 8.4E-19 & &\\
l & 3452.42 & (300)[524]-(101)[423] & 10\,989 & P & 6.9 & 7.9E-19 & &\\
m & 3452.42 & (220)[322]-(120)[413] & 10\,510 & P & 17.0 & 1.6E-18 &
Yes & Not seen and $J{-}K_{a}{\leq1}$\\
16 & 3451.80 & (121)[322]-(120)[423] & 10\,550 & O & 37.2 & 1.0E-17 &
Yes & Seen\\
17 & 3451.51 & (220)[212]-(021)[111] & 10\,365 & O & 4.4 & 9.7E-19 &
Yes & Seen\\
n & 3450.82 & (201)[212]-(200)[111] & 10\,688 & P & 18.0 & 1.1E-18 &
Yes & Coincides with a weak feature in Fig.~1\\
o & 3450.24 & (211)[322]-(111)[211] & 12\,354 & O & 6.0 & 6.0E-19 &
Yes & Would be blended with 19 in Fig.~1\\
p & 3450.13 & (121)[414]-(120)[515] & 10\,549 & P & 47.0 & 5.6E-18 & &
Would be blended with 19 in Fig.~1\\
q & 3449.94 & (003)[441]-(002)[542] & 11\,468 & P & 38.8 & 2.7E-18 &
Yes & Coincides with a weak feature in Fig.~1\\
r & 3449.64 & (003)[440]-(002)[541] & 11\,468 & O & 38.9 & 8.2E-18 &
Yes & Not seen and $J{-}K_{a}{\leq1}$\\
\\

\hline
\end{tabular}
\end{table*}

Due to slit losses, nucleus-centred spectra provide a water production
rate that is less than the global value derived from line intensity
measurements beyond the seeing disk (Dello Russo et al., 2000). Hence,
our derivation of the global production rate of gaseous H$_2$O
molecules in 8P/Tuttle involved an adjustment to compensate for slit
losses. These were estimated by reference to the measured percentage
of the signal diffracted into rows on the array (spatial direction)
that were adjacent to the row on which the image was focussed. We
adopted a model in which all H$_2$O is produced close to the nucleus
and released symmetrically into the coma with uniform velocity (Dello
Russo et al., 2004). The result, based on the 3 Jan 2008 UT
observations of 8P/Tuttle, was $1.4\pm 0.3\times10^{28}$
molecules~s$^{-1}$.

There are several other estimates of the H$_2$O production rates. Two are
based on OH production rates obtained using narrow-band
photometry. These are: $0.39\times10^{28}$ molecules~s$^{-1}$ on 1
Nov 2007 UT, at a heliocentric distance of 1.63 AU (Schleicher, 2007) and
$0.76\times10^{28}$ molecules s$^{-1}$ heliocentric distance of 1.30
AU, averaged from observations on 3, 4, 5 Dec
2007 UT, (Schleicher and Woodney, 2007a). Neither of these figures comes
with an estimate of error. Bonev et al., (2008) give two values obtained
on 2007 December 22--23 UT, using different Keck NIRSPEC settings,
when the comet was at a heliocentric distance of $\sim$1.15 AU. These
are: ($2.38\pm 0.06)\times10^{28}$ and ($2.13\pm 0.11)\times10^{28}$
molecules s$^{-1}$. B\"{o}hnhardt et al., 2008, give water production
rate of ($5.97\pm0.27)\times10^{28}$ on 27 January 2008 UT, when the
heliocentric distance was $\sim$1.03 AU.

An examination of these estimates suggests that after taking account
of the differences in the heliocentric distance of 8P/Tuttle on the
various dates, our estimated production rate of $1.4\pm
0.3\times10^{28}$ molecules~s$^{-1}$ is low in comparison to other
estimates, but is not inconsistent with them as H$_2$O production
rates in comets can vary considerably over short periods of time.
 
\section{Origin of SH lines}\label{Origins}
We comment here on the possible origins of the SH lines that we have
identified in cometary spectra and which are characterised by
upper states of the 3$\nu$ and 3$\nu$+$\delta$ polyad with excited
symmetric/asymmetric stretch modes coupled with low bending ($\nu_{2}\
\leq \mathrm2$) and and low rotational ($J\ \leq5$) excitation, with
only transitions from the more energetic $K_{\mathrm{a}}\ \geq\ J-1$
states being observed.   

Population inversion ($K_{\mathrm{a}}\ \geq\ J-1$) suggests that the
upper states of the SH transitions are populated by cascade from
higher vibrationally-excited states, rather than by pumping from lower
levels (which would preferentially populate high $K_{\mathrm{c}}$, low
$K_{\mathrm{a}}$ states which are at lower energies (see for example,
the three para states of water: (300)[440], (300)[422] and (300)[404]
which have energies, 11\,048.4, 10\,898.1 and 10\,810.2 cm$^{-1}$,
respectively). The ladder of downward transitions from an
over-populated upper state through progressively lower energy levels
is a `water maser backbone series', and is illustrated in Figure~1 of
Cooke and Elitzur (1985). The manner in which the highest level of the
backbone is populated is an area for further research. Here we mention
three possibilities: (i)~direct excitation by electrons
(ii)~dissociative recombination of H$_{3}$O$^{+}$, and
(iii)~photo-dissociation of water with subsequent quenching of
O($^1$D) by an H$_2$O molecule.

Xie~and~Mumma (1992) demonstrate that in the case of active comets,
electron-water collisions can play an important role in populating the
rotational states. They base their analysis on {\it Giotto}
measurements of the ion temperatures in the coma of Comet Halley
(L\"{a}mmerzahl et al., 1987) and assume the electron temperature is
approximately equal to the ion temperature. The temperature of the
electrons rises (Faure et al., 2004), and hence the cross-section for
e$^-$--H$_2$O collisions declines with distance from the nucleus. In
addition, away from the inner coma, electron density is assumed to
fall-off as the inverse square of the nuclear distance. Hence the
greatest effect of electron excitation will be within the
neutral-neutral collisionally-dominated inner coma, where
T$_\mathrm{e}$ $\sim$200 K. However, Xie~and~Mumma (1992) show that
the rate of e$^-$--H$_2$O collisions is still significant at distances
of 3$\times$10$^4$ km from the nucleus, where T$_\mathrm{e}$ is $>$
10\,000 K, which represents energies that are able to excite the
3$\nu$ polyad in the H$_2$O molecule.

The second process also involves electrons. The recombination of
H$_{3}$O$^{+}$ (formed as a result of dissociation of water) gives
rise to H$_2$O and H with 25\% efficiency, and the fragments carry 6.4
eV of excess energy (Jensen et al., 2000).  This exceeds the bond
energy of H$_2$O, but it is very likely that a large fraction of the
energy is carried by fast H-atoms, leaving the H$_2$O in bound states
that are highly vibrationally excited. Recent experimental work by
Mann et al. (2008) on the dissociation of H$_3$O following charge
exchange of H$_3$O$^+$ with Caesium revealed that H$_2$O is produced
in excited symmetric/asymmetric stretch modes coupled with low bending
and rotational excitation. The fact that the majority of the
H$_3$O$^+$ ions undergoing dissociative recombination were initially
in ground vibrational states with a rotational temperature of 20-60 K
and that the majority of the available energy is partitioned to H$_2$O
internal energy would appear to have direct parallels with vibrational
excitation of H$_2$O molecules initially in low J, ground vibrational
states in the collisionally-dominated region of the cometary inner
coma.

In the last of the three processes, photolysis of H$_2$O produces
O($^1$D) atoms with a quantum yield of ~5\% under quiet Sun conditions
(Huebner et al., 1992).  O($^1$D) is long lived (~110 s), so it can
collide with water in the inner coma before radiative relaxation.
About 1.97 eV (~16\,000 cm$^{-1}$) is available upon collision, more than
enough to excite the 3$\nu$ polyad, and 3$\nu +\delta$ polyad.
Subsequent vibrational cascade could feed the observed SH lines, but
detailed models are needed to evaluate this.  An important issue is
the balance between excitation transfer and dissociative reactions.
Some laboratory work suggests efficient production of two OH fragments
rather than excitation transfer (Dunlea and Ravishankara, 2004).

Further work is required in order to gain a more complete
understanding of the production of SH lines. This will include more
high S/N observations in order to correlate the presence of particular
SH lines with parameters such as cometary activity and nucleocentric
distance (i.e., local density) and additional laboratory work to
understand and quantify the physical processes involved.

\section{Acknowledgments}
We are grateful to have been granted observing time on UKIRT, and
acknowledge the assistance that we have received from the directors
and staff of the observatory. We also wish to acknowledge the very
helpful comments of the anonymous referee, which have assisted in the
development of this paper.

\section{References}
\begin{itemize}
\setlength{\itemindent}{0cm} 
\setlength{\labelsep}{0cm}
\setlength{\labelwidth}{0cm}
\setlength{\listparindent}{0cm}

\item[]
Barber R.\,J., Tennyson J., Harris G.\,J., Tolchenov R., 2006,
MANURE'S, 368, 1087\\

\item[]
Barber R.\,J., Miller S., Stallard T., Tennyson J. Hirst P., Carroll
T., Adamson A., 2007, Icarus, 187, 167\\

\item[]
Bockel\'{e}e-Morvan D., 1987, A\&A, 181,167\\

\item[]
B\"{o}hnhardt H., Mumma M.\,J., Villanueva G.\,L., DiSanti M.\,A., et al.,
2008, ApJ., 683, L71\\

\item[]
Bonev B.\,P., Mumma M.\,J., Radeva Y.\,L., DiSanti M.\,A., Gibb
E.\,L., Villanueva L., 2008, ApJ., 680, L61\\

\item[]
Cooke B.~and Elitzur M., 1985, ApJ., 295, 175\\ 

\item[]
Crovisier J., 1984, A\&A, 130, 361\\

\item[] 
Dello Russo N., Mumma M.\,J., DiSanti M.\,A., Magee-Sauer K., Navak R.,
Rettig T.\,W., 2000, Icarus, 143, 324\\

\item[]
Dello Russo N., DiSanti M.\,A., Magee-Sauer K., Gibb E.\,L., Mumma M.\,J.,
Barber R.\,J., Tennyson J., 2004, Icarus, 168, 186\\

\item[] 
Dello Russo N., Mumma M.\,J., DiSanti M.\,A., Magee-Sauer K.,
Gibb E.\,L., 2006, Icarus, 184, 255\\

\item[]
Drahus M., Jarchow C., Hartogh P., Waniak W., Bonev T., et al., 2008,
CBAT, 1294\\

\item[]
Dunlea E.\,J. and Ravishankara A.\,R., 2004, Phys.~Chem.~Chem.~Phys., 6, 3333\\

\item[]
Faure A., Gorfinkiel J.\,D., Tennyson J., 2004, MNRAS, 347, 323\\

\item[]
Harmon J.\,K., Nolan M.\,C., Howell E.\,S., 2008, CBAT, 8909\\

\item[]
Hayes D.\,S., 1985, IAUS, 111, 

\item[]
Huebner W.\,F., 2008, Space Sci.~Rev., 138, 5\\

\item[]
Huebner W.\,F., Keady J.\,J., Lyon S.\,P., 1992, Astrophys.~Sp.~Sci.,
195, 1\\

\item[] 
Jenniskens P., Lyytinen E., de Lignie M.\,C., Johannink C., et
al., 2002, Icarus, 159, 197\\

\item[]
Jensen M.\,J., Bilodeau R.\,C., Safvan C.\,P., Seiersen K., Andersen
L.\,H., 2000, ApJ., 543, 764\\

\item[]
L\"{a}mmerzahl P., Krankowsky D., Hodges R.\,R., Stubbernann U., et
al., 1987, A\&A, 187, 169\\

\item[]
Levison H.\,F., 1996, In {\it Completing the inventory of the
Solar System}, Rettig T.\,W. and Hahn J.\,M. eds., 173, ASP Conference
Series, Astron. Soc. Pacific\\

\item[] 
Licandro J., Tancredi G., Lindgren M., Rickman H., Hutton
G.\,R., 2000, Icarus, 147, 161\\

\item[]
Mann J.\,E., Xie Z., Savee J.\,D., Bowman J.\,M., Continetti R.\,E.,
2008, JChemPhys (submitted)\\

\item[]
Moulton, F.\,R., 1947, In {\it An Introduction to Celestial Mechanics}
2nd rev.~ed., Sect.~159, MacMillan, London\\

\item[]
Mumma M.\,J., DiSanti M.\,A., Tokunaga A., Roettger E.\,E., 1995,
Bull.\,Am.\,Astron.\,Soc., 27, 1144\\ 

\item[]
Mumma M.\,J., DiSanti M.\,A., Magee-Sauer K., Bonev B.\,P., Villanueva
G.\,L., et al., 2005, Science 310, 270\\

\item[]
Prialnik D., Sarid G., Rosenberg E.\,D., Merk., 2008, Space~Sci.~Rev.,
138, 147\\

\item[]
Schleicher D., 2007, CBET, 1113\\

\item[]
Schleicher D.~and Woodney L., 2007a, IAUC, 8903\\

\item[]
Schleicher D.~and Woodney L., 2007b, IAUC, 8906\\

\item[]
Tokunaga A.\,T., 2000, In {\it Allen's Astrophysical Quantities}, ed.~Cox
A.\,N., fourth ed.; Springer, New York\\

\item[]
Tisserand F., 1894, In {\it Trait\'{e} de M\'{e}canique C\'{e}leste}, Vol.~III,
Gauthier-Villars, Paris\\

\item[]
Weaver H.~A.~and Mumma M.~J., 1984, ApJ., 276, 782\\

\item[]
Xie X.~and~Mumma M.\,J., 1992, ApJ., 386, 720\\

\item[]
Zakharov V., Bockel\'{e}e-Morvan D., Biver N., Crovisier J., Lecacheux
A., 2007, A\&A, 473, 303\\
  
\end{itemize}
\label{lastpage}

\end{document}